# The Collaborate Calibration and Alignment of Button-type BPM's Offset

YUAN Jiandong(袁建东), MA Lizhen（马力祯）, ZHANG Bin（张斌）, YAO Junjie（姚俊杰）

（Institute of Modern Physics, Chinese Academy of Sciences, gansu, Lanzhou, 730000）

**Abstract:** Beam position monitor (BPM) can easily reinforce the handling of beam orbits and measure the absolute beam position [1]. Its data can be used to optimize and correct beam in both first turn and closed orbit mode. In order to set the absolute center position of Button-type BPM, and formulate the offset between mechanic and electronic center precisely, we mounted BPM together with solenoid on a vertical rotated test-bench when its calibration takes out, and developed "transform" software to calculate the offset. This paper describes the method and process of collaborate calibration: the assembly and alignment of BPM itself on the designed work-bench; the mechanic calibration of bundle BPM-Solenoid, and the alignment of mechanic to the wire center used by Laser Tracker and Portable coordinate measurement machine (CMM) jointly; the connection of coaxial cable and read-out for electronics; the electronic calibration of bundle BPM-Solenoid. Form the above four steps, the author analyses the error sources, measures and subtracts the offset values, and formulate the position calculation formula of offset. At last, the residual offset is nearly± 20μm. The total accuracy of calibration is 0.1mm.

## Introduction

BPM is the important component of accelerator diagnosis system, constituted by front-end detect electrode and after-end signal processing system. It measurements cover trajectory, first turn, equilibrium, turn-by-turn, beam based alignment and orbit stability. There are three types BPM: button-type; stripe line-type and shaped/cavity-type. Different from the conventional methods, this paper research discusses details of collaborate calibration method and error sources for button-type BPM.

Button-type BPM uses four capacity-coupling electrodes (called "button"), aims to measure the transverse and vertical beam position. A drawing of the monitor section is shown in Fig.1. It is a Button-type BPM made of stainless steel (SUS304) with rotational symmetry. A traversal length 40 mm was chosen, so that it can be installed into limited spaces of the Test CryMoudle6 in the ADS Injector ∏ beam line. A 44-mm-long bellows is attached to one side of the monitor. The monitors are fixed at the front-end of the solenoid magnet, and the center of the monitor is precisely aligned with that of the solenoid using 8 precise benchmark targets. The electrode's inner diameter was 11 mm. The other end of BPM is connected to a Half Wave Resonator (HWR), which is working in temperature about 4K.

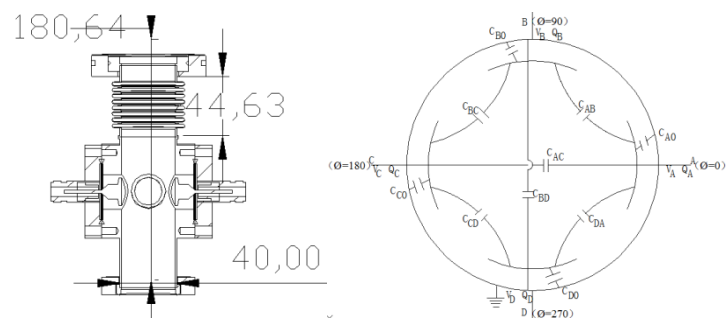

Fig.1 BPM Section drawing    Fig.2 Equivalent capacity coupling model

## 1 BPM principles and Characterization

According to the equivalent capacity coupling model [2] (as shown in Fig.2), A large angular width generates a larger electromagnetic coupling between the neighboring electrodes through equivalent capacitors but does not preserve a good linearity. Large angular widths are only determined by the geometrical configuration of the BPM electrodes and chamber radius. Bad linearity has a low sensitivity. The electronics of this paper is Libera Brilliance's difference-sum ratio ($\Delta/\Sigma$) [3]. The formula used to calculate the horizontal X and vertical Y absolute positions can be formulated as follows:

$$X = K_x \frac{V_A - V_C}{V_A + V_C} + X_{offset}$$

$$Y = K_y \frac{V_B - V_D}{V_B + V_D} + Y_{offset} \quad (1)$$

Where: Kx and Ky are the geometric coefficients of the BPM, VA, VB, VC and VD are the voltages on the corresponding electrodes scaled with the appropriate calibration coefficients, and $X_{offset}$, $Y_{offset}$ are the sum of mechanic, electronic and residual offset between mechanic and electronic center.

## 2 The process of calibration and alignment

Fig.3 shows a photograph of a rotated work-bench for the bundle BPM-Solenoid. A three-dimension adjustable stage and rotated platform are placed on a work-bench. Their installation error is less than 50 um. Every BPM is mounted on a calibrated Solenoid. The stage can move in a three-dimensional plane perpendicular to each other by using a stepping motor, which communicates with a personal computer. The absolute position of the BPM center is set both by mechanic and electronic method.

Because of the strictly requirements of symmetry in the four electrodes of button BPM, We used torque wrench and CMM to assembly and install the BPM itself and solenoid.

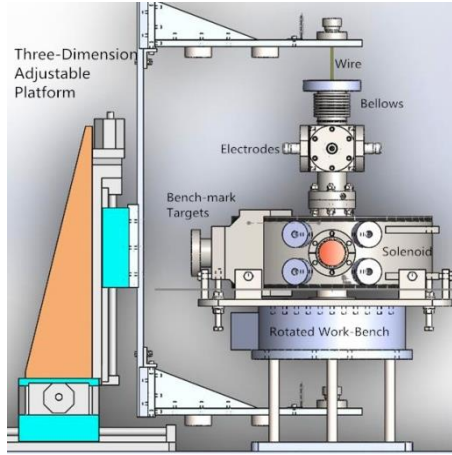 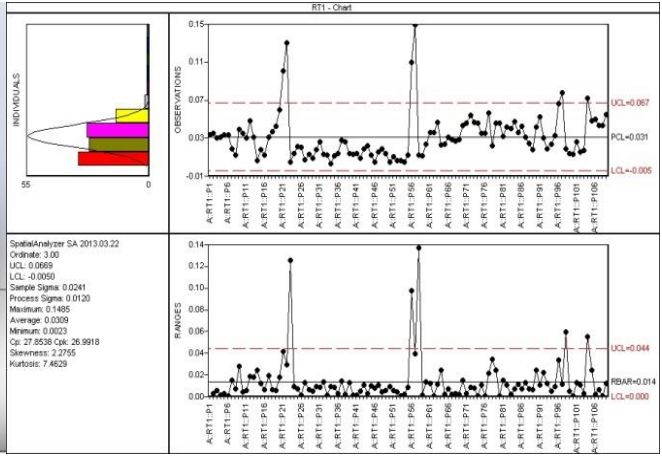

Fig.3 Rotated work-bench    Fig.4 Best fit process

### 2.1 The mechanic calibration of bundled BPM-Solenoid

By the use of CMM, we build the precise coordinate frame, which its origin is the mechanic center of bundle BPM-Solenoid, and Y axle is pure vertical, Z axle is along the beam direction.

Electronic calibration needs high performance of rotation symmetry, so that the center of mechanic should be aligning to the center of rotated platform before it takes out. Therefore, the offset between mechanic and rotation center need to be tracker measure when the electronic calibration takes out. Fig. 4 shows the best fit process of one of five bundles BPM-solenoid. Table1. shows the last result of adjustment.

Table 1. Offset values

|    | $DX_{electronic(mm)}$ | $K_x$  | $DX_{mechanic(mm)}$ | $DX_{residual(um)}$ |
|----|-----------------------|--------|---------------------|---------------------|
| 1# | -0.35                 | 1.0807 | -0.34               | -2                  |
| 2# | -0.14                 | 1.0846 | -0.23               | 4                   |
| 3# | -0.11                 | 1.0834 | -0.17               | 18                  |
| 4# | 0.15                  | 1.0868 | 0.15                | 5                   |
| 5# | 0.32                  | 1.0868 | 0.48                | -2                  |
|    | $DY_{electronic(mm)}$ | $K_y$  | $DY_{mechanic(mm)}$ | $DY_{residual(um)}$ |
| 1# | -0.49                 | 1.0836 | -0.59               | -5                  |
| 2# | -0.22                 | 1.083  | -0.21               | 1                   |
| 3# | 0.16                  | 1.0858 | 0.07                | 10                  |
| 4# | -0.28                 | 1.0856 | -0.39               | 0                   |
| 5# | 0.12                  | 1.0846 | 0.26                | 2                   |

### 2.2 The electronic calibration of bundled BPM-Solenoid

The electronic calibration is performed in two steps: first, for the determination of the intrinsic offset of the bundle BPM-Solenoid block itself. The intrinsic offset ($\Delta_{x0}$, $\Delta_{y0}$) is a constant. The second step consists cable and electronics offset ($\Delta_{xc}$, $\Delta_{yc}$). A coaxial cable is installed to connect the button-type BPM pick-up electrodes to the readout electronics. The cables and the electronic offset, as the input electronics is a difference-sum amplifier, are also strongly dependent on the input power and beam intensity. Fig.5 shows the calculate and survey process of the electronic offset. The total offset ($X_{electronic}$, $Y_{electronic}$) is the sum of above two values. Fig.6 Shows Calibration process of $K_x$ and $K_y$. These coefficients were determined by the coordinates of the probes and were used to process the output signals. The resolution of the electronics noise is proportional to k but the coefficient is unknown and varies from BPM to BPM[4].

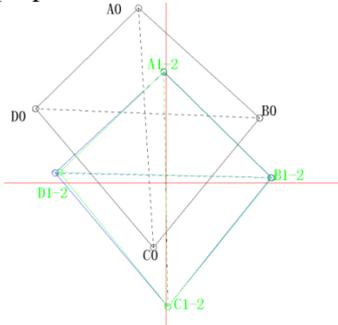 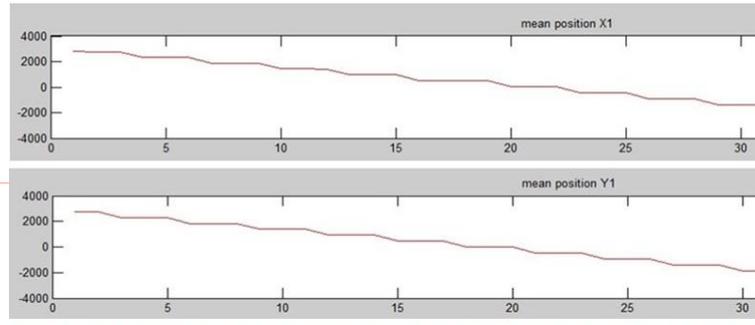

Fig.5 Electronic offset　　　　　　　　Fig.6 $K_x$-$K_y$

### 3 The error sources analysis

Intrinsic offset is due to unsymmetrical positions of the electrodes and relative difference of the buttons capacitance. The intrinsic offset is a constant. It comes from the assembly and alignment of BPM itself on the designed work-bench. Cable and electronics offset depends on the cable bends and therefore on their installation on the accelerator site. Cables and electronics calibration was performed before and after the installation on site without the beam.

The mechanic offset comes from the instrument and manual operation. The error of benchmark target's Manufacture is inevitable[5]. To guarantee the working accuracy, the future work is to improve the process tech.

### Conclusion

1 Fig.5 shows the electronic offset which contain the intrinsic and cable offset. It is a vector,

which its component at each axis indicators difference values of the electronic offset.

2 The actual residual offset can be seen in table 1;the nominal offset can be calculate form the Table1 and Formula 2：

$$X_{offset} = X_{electronic} + X_{mechanic} + X_{residual}$$
$$Y_{offset} = Y_{electronic} + Y_{mechanic} + Y_{residual} \quad (2)$$

The difference between nominal and actual offset also can be calculate. According to the Bessel formula, the mean square error of the difference is as follows:

$$\sigma_{\Delta x} = \sqrt{(0.01^2 + 0.09^2 + 0.06^2 + 0 + 0.16^2)/5} \approx 0.09 mm$$
$$\sigma_{\Delta y} = \sqrt{(0.1^2 + 0.01^2 + 0.09^2 + 0.11^2 + 0.14^2)/5} \approx 0.1 mm$$

It means that total accuracy of calibration is 0.1mm.